\documentclass[pra,twocolumn,graphicx]{revtex4}
\usepackage{amssymb}
\usepackage{epsfig,amsmath}

\begin{document}

\title{Influence of energetically close orbitals on molecular high-order
harmonic generation}
\author{C. Figueira de Morisson Faria and B. B. Augstein}
\affiliation{Department of Physics and Astronomy, University College London, Gower
Street, London WC1E 6BT, United Kingdom}
\date{\today}

\begin{abstract}
We investigate the contributions from the $3\sigma _{g}$ and $1\pi _{u}$ and
molecular orbitals in high-order harmonic generation in $N_{2}$, with
particular emphasis on quantum-interference effects. We consider both the
physical processes in which the electron is freed and returns to the same
orbital, and those in which it is ionized off one orbital and recombines
with the other. We show that the quantum-interference patterns observed in
the high-order harmonic spectra are predominantly determined by the $3\sigma
_{g}$ orbital. This holds both for the situation in which only the $1\pi
_{ux}$ orbital is considered, and the dynamics of the electron is restricted
to the plane $p_{x}p_{z},$ or in the full three-dimensional case, if the
azimuthal angle is integrated over and the degeneracy of $1\pi _{u}$ is
taken into account.
\end{abstract}

\maketitle

\section{Introduction}

In the past few years, high-order harmonic generation (HHG) has been
extensively studied as a tool for attosecond imaging. In particular
the possibility of bound-state reconstruction \cite{Itatani}, the
attosecond probing of dynamic processes in molecules \cite{attomol}
and quantum interference effects \cite{interfexp} has attracted a
great deal of attention. This is a consequence of the fact that HHG
is the result of the recombination of an electron, freed by
tunneling or multiphoton ionization at an instant $t^{\prime }$,
with its parent molecule at a later instant $t$ \cite{tstep}. Since,
in principle, the electron may recombine with more than one center,
high-harmonic emission at spatially separated sites takes place.
Hence, information about the structure of the molecule in question
is hidden in the HHG spectrum. In particular for diatomic molecules,
this can be thought of as a microscopic counterpart of the
double-slit experiment, in which maxima and minima arise due to the
two-center interference \cite{doubleslit}.

In many studies so far, it has been assumed that the electron is
released from the highest-occupied molecular orbital
(HOMO)\cite{MadsenN2,DM2009,DM2006,UsachenkoN2ion,MoreN2,LinN2,McFar2008}.
This, however, has been disputed in recent investigations, in which
it was shown that multielectron effects and the quantum interference
of different ionization channels may play an important role
\cite{Multielectron}. Such effects may constitute a serious obstacle
towards ultra-fast molecular imaging. Apart from that, even if only
the HOMO is considered as the initial state of the ionized electron,
in many cases its degeneracy has a considerable influence on the HHG
spectra \cite{degeneracy}.

In this paper, we investigate the influence of different molecular
orbitals on the high-order harmonic spectra of diatomic nitrogen
$(N_{2})$. High-order harmonic generation
\cite{MadsenN2,DM2009,Itatani,Multielectron,MoreN2,LinN2} and
above-threshold ionization \cite{MadsenN2,UsachenkoN2ion,DM2006} in
$N_{2}$ have been extensively investigated in the literature, as,
due to its large mass, its vibrational degrees of freedom do not
play a very important role and may be ignored to first
approximation. In fact, it has been shown that, whereas for lighter
species, vibration may lead to a considerable blurring of the
two-center interference patterns, and a reduction in the
high-harmonic or photoelectron yield, for molecular nitrogen such
effects are not significant \cite{MadsenN2}.

Furthermore, in $N_{2},$ the HOMO and the HOMO-1 orbitals are
energetically very close. This has several consequences. First,
since they possess opposite parity, one expects a strong coupling
between them. Second, since the tunneling probability is related to
the bound-state energy, the processes in which the electron starts
in the HOMO and in the HOMO-1 are comparable. Third, the electron
may also leave from one orbital and recombine with the other, and,
quantum mechanically, the transition amplitude related to all
physical processes involved will interfere. The influence of the
HOMO-1 in the high-harmonic spectra of $N_2$ has been recently
observed \cite{McFar2008}.

One should note, however, that, for $N_2$, the HOMO and the HOMO-1 exhibit
very distinct shapes and symmetry. In fact, the former is a $3\sigma _{g}$
orbital and the latter a $1\pi _{u}$ orbital. Therefore, they are expected
to behave differently as the alignment angle of the molecule with regard to
the laser-field polarization is varied. Apart from that, the $1\pi_{u}$
orbital is doubly degenerate.

In our investigations, we employ the strong-field approximation \cite{hhgsfa}%
, and saddle-point methods. The transition amplitudes obtained
within this framework can be related to the classical orbits of an
electron in a time-dependent field, and, yet, they retain
information on the quantum interference between the possible
physical processes \cite{orbitshhg}. Throughout, we employ the
length gauge. Even though there is considerable debate about which
gauge to employ, and the length gauge SFA leads to potential-energy
shifts whose meaning are not clear \cite{gauge}, it has been
recently shown that the two-center interference patterns are absent
in SFA computations of the high-harmonic spectra using the
velocity-gauge \cite{F2007,SSY2007,DM2009}.

This paper is organized as follows. In Sec. \ref{transampl}, we
provide the SFA transition amplitudes for the physical processes
involved, for an exponential basis set involving Slater-type
orbitals, and for a split-valence, gaussian basis set. Subsequently,
in Sec. \ref{results}, we compare the high-harmonic spectra obtained
using both basis sets (Sec. \ref{basisset}),
and investigate quantum-interference effects between the $3\sigma_g$ and $%
1\pi_u$ orbital (Sec. \ref{interference}). Finally, in Sec. \ref{concl} we
summarize the paper and state our main conclusions.

\section{Transition amplitudes}

\label{transampl}

Below we provide the HHG transition amplitudes, within the strong-field
approximation. We base our approach on the explicit expression in Ref. \cite%
{hhgsfa}, and employ atomic units throughout.

The HHG amplitude is generalized to the case in which the active electron is
initially in a coherent superposition of the $3\sigma _{g}$ and the $%
1\pi_{u} $ orbitals.

Explicitly,
\begin{equation*}
\left\vert \psi _{0}\right\rangle =C_{3\sigma _{g}}\left\vert 3\sigma
_{g}\right\rangle +C_{1\pi _{ux}}\left\vert 1\pi _{ux}\right\rangle +C_{1\pi
_{uy}}\left\vert 1\pi _{uy}\right\rangle ,
\end{equation*}%
where the coefficients $C_{3\sigma _{g}}$, $C_{1\pi _{ux}}$ and $C_{1\pi
_{uy}}$ give the weights of each state. One should note that the orbitals $%
1\pi _{ux}$ and $1\pi _{uy}$ are degenerate, and possess the energy $E_{1\pi
u}$. In the present model, we will neglect the processes in which the
electron, immediately before ionization, is excited from the $1\pi _{u}$
state to $3\sigma _{g},$ and, upon recombination, decays from $3\sigma _{g}$
to $1\pi _{u}.$

Under these assumptions, the overall transition amplitude will be the sum%
\begin{equation}
M=\sum\limits_{j,\nu }M_{j\nu }+c.c.
\end{equation}%
of nine terms. Explicitly,%
\begin{eqnarray}
M_{11} &=&\hspace*{-0.1cm}-i\left\vert C_{3\sigma _{g}}\right\vert ^{2}%
\hspace*{-0.1cm}\int_{-\infty }^{+\infty }\hspace*{-0.1cm}dt\hspace*{-0.1cm}%
\int\limits_{-\infty }^{t}\hspace*{-0.1cm}dt^{\prime }\int d^{3}pd_{z}^{\ast
(3\sigma _{g})}(\mathbf{p}+\mathbf{A}(t))  \notag \\
&&\times d_{z}^{(3\sigma _{g})}(\mathbf{p}+\mathbf{A}(t^{\prime }))\exp
[iS_{11}(t,t^{\prime },\mathbf{p})],  \label{M11}
\end{eqnarray}%
\begin{eqnarray}
M_{12} &=&\hspace*{-0.1cm}-iC_{3\sigma _{g}}^{\ast }C_{1\pi _{ux}}\hspace*{%
-0.1cm}\int_{-\infty }^{+\infty }\hspace*{-0.1cm}dt\hspace*{-0.1cm}%
\int\limits_{-\infty }^{t}\hspace*{-0.1cm}dt^{\prime }\int d^{3}pd_{z}^{\ast
(3\sigma _{g})}(\mathbf{p}+\mathbf{A}(t))  \notag \\
&&\times d_{z}^{(1\pi _{ux})}(\mathbf{p}+\mathbf{A}(t^{\prime }))\exp
[iS_{12}(t,t^{\prime },\mathbf{p})],  \label{M12}
\end{eqnarray}%
\begin{eqnarray}
M_{13} &=&\hspace*{-0.1cm}-iC_{3\sigma _{g}}^{\ast }C_{1\pi _{uy}}\hspace*{%
-0.1cm}\int_{-\infty }^{+\infty }\hspace*{-0.1cm}dt\hspace*{-0.1cm}%
\int\limits_{-\infty }^{t}\hspace*{-0.1cm}dt^{\prime }\int d^{3}pd_{z}^{\ast
(3\sigma _{g})}(\mathbf{p}+\mathbf{A}(t))  \notag \\
&&\times d_{z}^{(1\pi _{uy})}(\mathbf{p}+\mathbf{A}(t^{\prime }))\exp
[iS_{13}(t,t^{\prime },\mathbf{p})],
\end{eqnarray}%
\begin{eqnarray}
M_{21} &=&\hspace*{-0.1cm}-iC_{3\sigma _{g}}C_{1\pi _{ux}}^{\ast }\hspace*{%
-0.1cm}\int_{-\infty }^{+\infty }\hspace*{-0.1cm}dt\hspace*{-0.1cm}%
\int\limits_{-\infty }^{t}\hspace*{-0.1cm}dt^{\prime }\int d^{3}pd_{z}^{\ast
(1\pi _{ux})}(\mathbf{p}+\mathbf{A}(t))  \notag \\
&&\times d_{z}^{(3\sigma _{g})}(\mathbf{p}+\mathbf{A}(t^{\prime }))\exp
[iS_{21}(t,t^{\prime },\mathbf{p})],  \label{M21}
\end{eqnarray}%
\begin{eqnarray}
M_{22} &=&\hspace*{-0.1cm}-i\left\vert C_{1\pi _{ux}}\right\vert ^{2}%
\hspace*{-0.1cm}\int_{-\infty }^{+\infty }\hspace*{-0.1cm}dt\hspace*{-0.1cm}%
\int\limits_{-\infty }^{t}\hspace*{-0.1cm}dt^{\prime }\int d^{3}pd_{z}^{\ast
(1\pi _{ux})}(\mathbf{p}+\mathbf{A}(t))  \notag \\
&&\times d_{z}^{(1\pi _{ux})}(\mathbf{p}+\mathbf{A}(t^{\prime }))\exp
[iS_{22}(t,t^{\prime },\mathbf{p})],  \label{M22}
\end{eqnarray}%
\begin{eqnarray}
M_{23} &=&\hspace*{-0.1cm}-iC_{1\pi _{ux}}^{\ast }C_{1\pi _{uy}}\hspace*{%
-0.1cm}\int_{-\infty }^{+\infty }\hspace*{-0.1cm}dt\hspace*{-0.1cm}%
\int\limits_{-\infty }^{t}\hspace*{-0.1cm}dt^{\prime }\int d^{3}pd_{z}^{\ast
(1\pi _{ux})}(\mathbf{p}+\mathbf{A}(t))  \notag \\
&&\times d_{z}^{(1\pi _{uy})}(\mathbf{p}+\mathbf{A}(t^{\prime }))\exp
[iS_{23}(t,t^{\prime },\mathbf{p})],
\end{eqnarray}%
\begin{eqnarray}
M_{31} &=&\hspace*{-0.1cm}-iC_{3\sigma _{g}}C_{1\pi _{uy}}^{\ast }\hspace*{%
-0.1cm}\int_{-\infty }^{+\infty }\hspace*{-0.1cm}dt\hspace*{-0.1cm}%
\int\limits_{-\infty }^{t}\hspace*{-0.1cm}dt^{\prime }\int d^{3}pd_{z}^{\ast
(1\pi _{uy})}(\mathbf{p}+\mathbf{A}(t))  \notag \\
&&\times d_{z}^{(3\sigma _{g})}(\mathbf{p}+\mathbf{A}(t^{\prime }))\exp
[iS_{31}(t,t^{\prime },\mathbf{p})],
\end{eqnarray}%
\begin{eqnarray}
M_{32} &=&\hspace*{-0.1cm}-iC_{1\pi _{uy}}^{\ast }C_{1\pi _{ux}}\hspace*{%
-0.1cm}\int_{-\infty }^{+\infty }\hspace*{-0.1cm}dt\hspace*{-0.1cm}%
\int\limits_{-\infty }^{t}\hspace*{-0.1cm}dt^{\prime }\int d^{3}pd_{z}^{\ast
(1\pi _{uy})}(\mathbf{p}+\mathbf{A}(t))  \notag \\
&&\times d_{z}^{(1\pi _{ux})}(\mathbf{p}+\mathbf{A}(t^{\prime }))\exp
[iS_{32}(t,t^{\prime },\mathbf{p})],
\end{eqnarray}%
\begin{eqnarray}
M_{33} &=&\hspace*{-0.1cm}-i\left\vert C_{1\pi _{uy}}\right\vert ^{2}%
\hspace*{-0.1cm}\int_{-\infty }^{+\infty }dt\hspace*{-0.1cm}%
\int\limits_{-\infty }^{t}\hspace*{-0.1cm}dt^{\prime }\hspace*{-0.1cm}\int
d^{3}pd_{z}^{\ast (1\pi _{uy})}(\mathbf{p}+\mathbf{A}(t))  \notag \\
&&\times d_{z}^{(1\pi _{uy})}(\mathbf{p}+\mathbf{A}(t^{\prime }))\exp
[iS_{33}(t,t^{\prime },\mathbf{p})],
\end{eqnarray}%
where $d_{z}^{(3\sigma _{g})}(\mathbf{p})=\left\langle \mathbf{p}\right\vert
\mathbf{r}.\hat{e}_{z}\left\vert 3\sigma _{g}\right\rangle $ and $%
d_{z}^{(1\pi _{u\xi })}(\mathbf{p})=\left\langle \mathbf{p}\right\vert
\mathbf{r}.\hat{e}_{z}\left\vert 1\pi _{u\xi }\right\rangle ,(\xi =x,y)$ are
the components of the dipole matrix elements related to $3\sigma _{g}$ and $%
1\pi _{u}$ along the field-polarization axis.

In the above-stated equations, one may distinguish two types of
contributions. The amplitudes $M_{jj}$ correspond to the processes in which
the electron leaves a specific orbital, reaches a Volkov state $\left\vert%
\mathbf{p}+\mathbf{A}(t^{\prime})\right\rangle$, propagates in the continuum
from $t^{\prime}$ to $t$, and recombines from a Volkov state $\left\vert%
\mathbf{p}+\mathbf{A}(t)\right\rangle$ to the same orbital it left from. The
amplitudes $M_{j\nu },$ $j\neq \nu ,$ on the other hand, give the processes
in which the electron leaves from one orbital and returns to another. The
corresponding actions read
\begin{equation}
S_{jj}(t,t^{\prime },\mathbf{p})=S(t,t^{\prime },\mathbf{p})-E_{\alpha
}(t-t^{\prime }),
\end{equation}%
and
\begin{equation}
S_{j\nu }(t,t^{\prime },\mathbf{p})=S(t,t^{\prime },\mathbf{p})-(E_{\alpha
}t-E_{\beta }t^{\prime }),\ j\neq \nu ,
\end{equation}%
respectively, with%
\begin{equation}
S(t,t^{\prime },\mathbf{p})=\Omega t-\frac{1}{2}\int_{t^{\prime }}^{t}d\tau %
\left[ \mathbf{p}+\mathbf{A}(\tau )\right] ^{2}.
\end{equation}
In the above-stated equations, $E_{\alpha },E_{\beta }$ refer to the
bound-state energies, $t$ to the recombination time, $t^{\prime }$ to the
start time and $\mathbf{p}$ the intermediate momentum. For \thinspace $%
S_{11},\alpha =3\sigma _{g},$ while for \thinspace $S_{22}$ and $%
S_{33},\alpha =1\pi _{u}$. For $S_{12}(t,t^{\prime },\mathbf{p}),$ $%
E_{\alpha }=E_{3\sigma _{g}}$ and $E_{\beta }=E_{1\pi _{u}},$whilst, for $%
S_{21}(t,t^{\prime },\mathbf{p}),$ the situation is reversed, i.e., $%
E_{\alpha }=E_{_{1\pi _{u}}}$ and $E_{\beta }=E_{3\sigma _{g}}$. Note that,
due to the fact that the $1\pi _{u}$ orbitals are degenerate, $%
S_{12}=S_{13}, $ $S_{21}=S_{31}$ and $S_{32}=S_{22}=S_{33}=S_{23}.$

We will compute the transition amplitude $M$ employing the stationary phase
approximation, i.e., we will look for values of $t,t^{\prime }$ and $\mathbf{%
p}$ that renders the actions in Eqs. (\ref{M11})-(\ref{M22})
stationary. Apart from considerably simplifying the computations
involved, this approach provides a physical interpretation of the
amplitudes $M_{j\nu }$ in terms of electron trajectories. We compute
the transition amplitudes employing a uniform saddle-point
approximation. Details on the specific method used can be found in
\cite{atiuni}.

\subsection{Saddle-point equations}

Differentiating $S_{11}$, $S_{22}$ and $S_{33}$ with respect to the
ionization time $t^{\prime }$ and the recombination time $t,$ we obtain the
saddle-point equations%
\begin{equation}
\frac{\left[ \mathbf{p}+\mathbf{A}(t^{\prime })\right] ^{2}}{2}+E_{\alpha }=0
\label{tunnelsame}
\end{equation}%
and
\begin{equation}
\frac{\left[ \mathbf{p}+\mathbf{A}(t)\right] ^{2}}{2}+E_{\alpha }=\Omega ,
\label{recsame}
\end{equation}%
where $\alpha =3\sigma _{g}$ for $S_{11}(t,t^{\prime },\mathbf{p})$ and $%
\alpha =1\pi _{u}$ for $S_{22}$ and $S_{33}.$ Physically, Eq. (\ref%
{tunnelsame}) gives the conservation of energy at the instant of tunneling,
and Eq. (\ref{recsame}) expresses the fact that the electron recombines to
the \textit{same} state (either $\left\vert 3\sigma _{g}\right\rangle $ or $%
\left\vert 1\pi _{u}\right\rangle $), releasing its kinetic energy upon
return in form of a high-order harmonic of frequency $\Omega $. Finally, the
condition $\partial S_{jj}/\partial \mathbf{p}=\mathbf{0}$ yields

\begin{equation}
\int_{t^{\prime }}^{t}d\tau \left[ \mathbf{p}+\mathbf{A}(\tau )\right] =%
\mathbf{0}  \label{return}
\end{equation}%
Eq. (\ref{return}) constrains the intermediate momentum of the electron, so
that it returns to the site of its release. In the present model, this site
is taken as the origin of our coordinate system, and is the geometric center
of the molecule. Summarizing, the saddle-point equations (\ref{tunnelsame})-(%
\ref{return}) are related to the physical picture of an electron ionizing
from either the HOMO or the HOMO-1 in $N_{2}$ and returning to the same
state.

The remaining actions $S_{j\nu }(t,t^{\prime },\mathbf{p}),$ for $j\neq \nu,
$ lead to the saddle-point equations

\begin{equation}
\frac{\left[ \mathbf{p}+\mathbf{A}(t^{\prime })\right] ^{2}}{2}+E_{\beta }=0
\label{tunneldiff}
\end{equation}%
and
\begin{equation}
\frac{\left[ \mathbf{p}+\mathbf{A}(t)\right] ^{2}}{2}+E_{\alpha }=\Omega ,
\end{equation}%
which indicate that the electron has left from one state and recombined with
the other. For $S_{12}(t,t^{\prime },\mathbf{p})$ and $S_{13}(t,t^{\prime },%
\mathbf{p}),$ $E_{\alpha }=E_{3\sigma _{g}}$ and $E_{\beta }=E_{1\pi _{u}},$%
while, for $S_{21}(t,t^{\prime },\mathbf{p})$ and $S_{31}(t,t^{\prime },%
\mathbf{p}),$ the situation is reversed, i.e., $E_{\alpha }=E_{_{1\pi _{u}}}$
and $E_{\beta }=E_{3\sigma _{g}}$. For the remaining terms, $E_{\alpha
}=E_{\beta }=E_{_{1\pi _{u}}}$ so that Eqs. (\ref{tunnelsame}) and (\ref%
{recsame}) are recovered. Physically, this corresponds to the situation in
which the electron leaves $\left\vert 1\pi _{ux}\right\rangle $ and returns
to $\left\vert 1\pi _{uy}\right\rangle ,$ or vice-versa. The return
condition (\ref{return}) remains the same in this case.

\subsection{Orbital wavefunctions and dipole matrix elements}

Within the framework of the strong-field approximation, all structural
information about the molecule is embedded in the\ recombination prefactor $%
d_{z}^{(\Psi )}(\mathbf{p+A}(t))=\left\langle \mathbf{p+A}(t)\right\vert
\mathbf{r}.\hat{e}_{z}\left\vert \Psi \right\rangle ,$ with $\Psi =3\sigma
_{g}$ or $1\pi _{u}.$ In position space, this prefactor is given by
\begin{equation}
d_{z}^{(\Psi )}(\mathbf{p})=\frac{1}{(2\pi )^{3/2}}\int d^{3}r\mathbf{p}%
\cdot \hat{e}_{z}\exp [-i\mathbf{p}\cdot \mathbf{r}]\Psi (\mathbf{r}),
\end{equation}%
i.e., the component of $i\partial _{\mathbf{p}}\Psi (\mathbf{p})$ along the
laser-field polarization. In the following, we will construct \ the
momentum-space wavefunction $\Psi (\mathbf{p})$ for both orbitals. We will
consider the linear combination of atomic orbitals (LCAO) approximation and
frozen nuclei. This implies that the position-space wavefunction reads%
\begin{equation}
\Psi (\mathbf{r})=\sum_{a}\psi _{a}(\mathbf{r+R}/2)+(-1)^{l_{a}-m_{a}+%
\lambda _{a}}\psi _{a}(\mathbf{r-R}/2),  \label{LCAOposition}
\end{equation}%
where $\mathbf{R,}$ $l_{a}$ and $m_{a}$ denote the internuclear separation,
the orbital and magnetic quantum numbers, respectively$.$ For gerade and
ungerade symmetry, $\lambda _{a}=\left\vert m_{a}\right\vert $ and $\lambda
_{a}=\left\vert m_{a}\right\vert +1$, respectively.

Throughout, we will use the length form of the dipole operator and neglect
the terms growing linearly with the internuclear separation. Such terms are
artifacts and come from the lack of orthogonality between the Volkov states
and the field-free bound states. For a more complete discussion see \cite%
{dressedSFA,JMOCL2006,F2007}.

The wave functions $\psi _{a}(\mathbf{r})$ will be approximated by either
exponentially decaying, Slater-type orbitals or by a gaussian basis set. In
the former case,
\begin{equation}
\psi _{a}^{(HF)}(\mathbf{r})=\frac{c_{a}(2\zeta _{a})^{n_{a}+1/2}}{\sqrt{%
(2n_{a})!}}r^{n_{a}-1}e^{-\zeta _{a}r}Y_{l_{a}}^{m_{a}}(\theta ,\phi ),
\label{HFposition}
\end{equation}%
where $n_{a}$ refers to the principal quantum number, and, in the latter,
\begin{equation}
\psi _{a}^{(G)}(\mathbf{r})=\sum\limits_{j}\tilde{c}_{aj}\varphi _{j}^{(G)}(%
\mathbf{r})
\end{equation}%
with%
\begin{equation}
\varphi _{j}^{(G)}(\mathbf{r})=\sum_{\nu }b_{\nu }x^{\beta _{x}}y^{\beta
_{y}}z^{\beta _{z}}\exp [-\zeta _{\nu }r^{2}].
\end{equation}%
The coefficients $c_{a},$ $\tilde{c}_{aj}$ and $b_{\nu }$ and the exponents $%
\zeta $ are extracted either from quantum chemistry codes, or from existing
literature.

An exponential basis set has been recently employed in the
literature \cite{MadsenN2,DM2006,DM2009}, while the use of gaussians
is more widespread within the quantum chemistry community. In
particular, a gaussian basis set exhibits several advantages.

First, it allows an easier evaluation of the momentum-space
wavefunction, which will be a central ingredient for computing the
matrix elements $d_{z}^{(\Psi )}(\mathbf{p+A}(t)).$ Second, within
the SFA framework, for exponentially decaying states, the ionization
prefactor $d_{z}^{(\Psi )}(\mathbf{p+A}(t^{\prime }))$ exhibits a
singularity, according to the saddle-point equations (\ref{tunnelsame}) and (%
\ref{tunneldiff}). In previous work, we have eliminated this singularity by
incorporating the prefactor $d_{z}^{(\Psi )}(\mathbf{p+A}(t^{\prime }))$ in
the action, and found out that it did not play a considerable role \cite%
{singularity}. This singularity, however, is absent if gaussian
wavefunctions are taken. Finally, in Hartree Fock computations there is an
artifact that renders the $1\pi _{u}$ orbital more loosely bound than $%
3\sigma _{g}.$

Explicitly, the Fourier transform of \ Eq. (\ref{LCAOposition}) reads%
\begin{equation}
\Psi (\mathbf{p})=\sum_{a}\eta (l_{a},m_{a},\mathbf{p}+\mathbf{A}(t))\psi
_{a}(\mathbf{p}),  \label{pspaceHF}
\end{equation}%
with%
\begin{equation}
\eta (l_{a},m_{a},\mathbf{p})=\mathcal{C}_{+}\cos \left[ \frac{\mathbf{%
p\cdot R}}{2}\right] +i\mathcal{C}_{-}\sin \left[ \frac{\mathbf{p\cdot R}}{2}%
\right]
\end{equation}%
and%
\begin{equation}
\mathcal{C}_{\pm }=\pm 1+(-1)^{l_{a}-m_{a}+\lambda _{a}}.
\end{equation}%
A generalized interference condition, which takes into account the
structure of the orbitals in question, such as, for instance the
$s$-$p$ mixing in the $3\sigma _{g}$ orbital, can be inferred from
Eq. (\ref{pspaceHF}). Indeed,
if we consider $\vartheta =\arctan (i\mathcal{C}_{+}/\mathcal{C}_{-}),$ then%
\begin{equation}
\eta (l_{a},m_{a},\mathbf{p})=\sqrt{\mathcal{C}_{+}^{2}-\mathcal{C}_{-}^{2}}%
\sin [\vartheta +\mathbf{p\cdot R}/2].  \label{interfgeneral}
\end{equation}%
For $\eta (l_{a},m_{a},\mathbf{p}+\mathbf{A}(t))$, interference minima are
present if%
\begin{equation}
\vartheta +[\mathbf{p}+\mathbf{A}(t)]\cdot \mathbf{R}/2=\kappa \pi ,
\label{minima}
\end{equation}%
where $\kappa $ denotes an integer number. This interference
condition has been first derived in \cite{DM2009}.

For Slater-type orbitals the individual wavefunctions are given by%
\begin{eqnarray}
\psi _{a}^{(HF)}(\mathbf{p}) &=&\frac{(-ip)^{l_{a}}2^{n_{a}-l_{a}}\zeta
_{a}^{-(l_{a}+3/2)}}{\sqrt{(2n_{a})!}}\frac{\Gamma (2+l_{a}+n_{a})}{\Gamma
(3/2+n_{a})}  \notag \\
&&\times _{2}F_{1}(\alpha _{1},\alpha _{2},\alpha _{3},\alpha
_{4})Y_{l_{a}}^{m_{a}}(\theta _{p},\phi _{p}),  \label{PsiaHFsigm}
\end{eqnarray}%
and the arguments of the hypergeometric functions read $\alpha _{1}=$ $%
1+(l_{a}+n_{a})/2,$ $\alpha _{2}=$ $\alpha _{1}+1/2,$ $\alpha
_{3}=l_{a}+3/2,\alpha _{4}=-p^{2}/\zeta _{a}^{2}.$ The angles are given by $%
\theta _{p}=\cos ^{-1}(p_{z}/p)$ and $\phi _{p}=\tan ^{-1}(p_{y}/p_{x})$.

It is worth noticing that Eq. (\ref{PsiaHFsigm}) is mainly employed in the
description of $\sigma $ orbitals, since the spherical harmonics $%
Y_{l_{a}}^{m_{a}}(\theta _{p},\phi _{p})$ are real for $m_{a}=0.$
For $\pi $ orbitals, it makes physically more sense to employ real
spherical harmonics, which are linear combinations of
$Y_{l_{a}}^{m_{a}}(\theta _{p},\phi _{p})$ and
$Y_{l_{a}}^{-m_{a}}(\theta _{p},\phi _{p}).$ The explicit
expressions for the real spherical harmonics are provided in
\cite{DM2009}.

For a gaussian basis set the wavefunction $\psi _{a}(\mathbf{p})$ reads%
\begin{equation}
\psi _{a}^{(G)}(\mathbf{p})=\sum\limits_{j,\nu }\tilde{c}_{aj}b_{\nu }\tilde{%
\varphi}_{\nu }(\mathbf{p}),  \label{pspaceGAMESS}
\end{equation}%
with%
\begin{equation}
\tilde{\varphi}_{\nu }(\mathbf{p})=\prod\limits_{k}\tilde{\varphi}_{\nu
}(p_{k})  \label{Gaussian3D}
\end{equation}%
and $k=x,y,z.$ Explicitly,
\begin{equation}
\tilde{\varphi}_{\nu }(p_{k})=\frac{1}{2}\zeta ^{-\frac{\beta k}{2}-1}\left(
\chi (p_{k})+\Xi (p_{k})\right) ,  \label{Gaussian1}
\end{equation}%
$\tilde{\xi}$with%
\begin{eqnarray}
\chi (p_{k}) &=&i\left( -1+(-1)^{\beta _{k}}\right) p_{k}\Gamma \left( \xi
_{k}+\frac{1}{2}\right) \,  \notag \\
&&\times _{1}F_{1}\left( \xi _{k}+\frac{1}{2};\frac{3}{2};\tilde{\xi}%
_{k}\right)
\end{eqnarray}%
and%
\begin{equation}
\Xi (p_{k})=\left( 1+(-1)^{\beta _{k}}\right) \sqrt{\zeta _{\nu }}\Gamma
\left( \xi _{k}\right) \,_{1}F_{1}\left( \xi _{k};\frac{1}{2};\tilde{\xi}%
_{k}\right) .
\end{equation}%
The arguments of the Hypergeometric functions are denoted by $\xi
_{k}=(\beta _{k}+1)/2$ and $\tilde{\xi}_{k}=-p_{k}^{2}/(4\zeta _{\nu }).$ In
this work, we will be using $p$ and $s$ states, so that Eq. (\ref{Gaussian3D}%
) will reduce to
\begin{equation}
\tilde{\varphi}_{\nu }(\mathbf{p})=(-ip_{k})^{l_{a}}\frac{\pi ^{3/2}}{%
2^{l_{a}}\zeta _{\nu }^{3/2+l_{a}}}\exp [-p^{2}/(4\zeta _{\nu })].
\end{equation}%
Therein, $k=z,$ $k=x$ and $k=y$ for $\sigma ,$ $\pi _{x}$ and $\pi _{y}$
states, respectively. The return condition (\ref{return}) guarantees that
the momentum $\mathbf{p}$ and the external field are collinear. Hence, for a
linearly polarized field $\theta _{p}$ is equal to the alignment angle $%
\theta _{L}.$

\section{Harmonic spectra}

\label{results}

In the following, we will present the high-order harmonic spectra. We choose
the driving field as a linearly polarized monochromatic wave of frequency $%
\omega $ and amplitude $\omega A_{0}$ directed along the axis $z.$ Hence,
the corresponding vector potential is%
\begin{equation}
\mathbf{A}(t)=A_{0}\cos (\omega t)\hat{e}_{z}
\end{equation}
For all situations, we consider starting times $0<t^{\prime }<\omega
\pi $ confined to the first half cycle and the three shortest pairs
of orbits. For this particular field, using the saddle-point
equation (\ref{recsame}), the generalized interference condition
(\ref{minima}) may be expressed in terms
of the harmonic order $n$ as%
\begin{equation}
n=\frac{E_{\alpha }}{\omega }+\frac{2(\kappa -\vartheta )^{2}}{\omega
R^{2}\cos ^{2}\theta _{L}},
\end{equation}
where $E_{\alpha}$ is the absolute value of the bound-state energy
in question, $\kappa$ is an integer number, $\theta_L$ is the
alignment angle, $R$ is the internuclear distance and $\vartheta$ is
defined in Eq. (\ref{interfgeneral}).

\subsection{HOMO and HOMO-1 contributions}

\label{basisset} In this section, we will make an assessment of the
main differences encountered in the HHG spectra if the orbitals are
built employing a split valence, gaussian basis set, or
exponentially decaying, Slater-type orbitals. For that purpose, we
will concentrate on the recombination prefactor $d_{z}^{(\Psi
)}(\mathbf{p+A}(t))$
and assume that the ionization prefactor $d_{z}^{(\Psi )}(\mathbf{p+A}%
(t^{\prime }))$ is constant and unitary.

As a starting point, a direct comparison with the results reported in \cite%
{DM2009} for the $3\sigma _{g}$ orbital will be performed. This orbital is
known to exhibit a strong mixing between $s$ and $p$ states. Therefore, we
will address the question of how such a mixing influences the overall
interference patterns. In the context of the present article, this implies
that we will consider the transition amplitude $M_{11}$, for which the
electron leaves and recombines with the $3\sigma_{g}$ orbital.

\begin{figure}[ptb]
\noindent\hspace*{-0.5cm}\includegraphics[width=10.5cm]{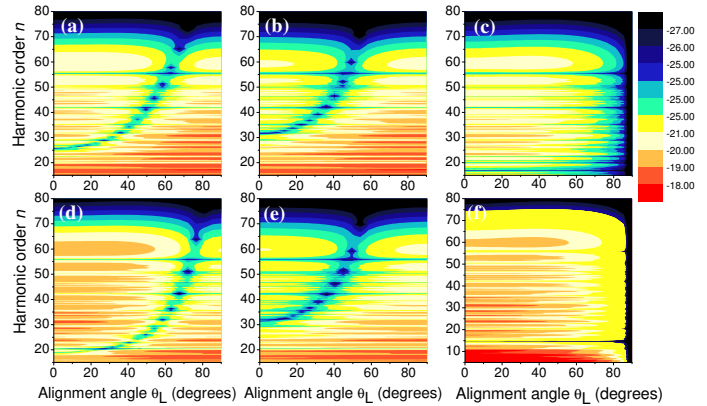}
\caption{High-order harmonic spectra for the HOMO in $N_2$ subject
to a linearly polarized laser field of frequency
$\protect\omega=0.057$ a.u. and intensity $I=4 \times
10^{14}\mathrm{W/cm}^2$, as a function of the alignment angle
$\protect\theta_L$ between the molecule and the field. The spectra
in the upper panels have been constructed using a gaussian basis set
and coefficients obtained from GAMESS-UK \protect\cite{GAMESS},
while those in the lower panels have been built using Slater-type
orbitals, and the coefficients in \protect\cite{Cade66}. From left
to right, we display the spectra from the full $3\protect\sigma_g$
orbital [panels (a) and (d)], the
contributions from the $s$ states [panels (b) and (e)], and those from the $%
p $ states [panels (c) and (f)]. The bound-state energy of the HOMO
and the equilibrium internuclear distance have been taken from the
respective computations. For the upper panels,
$E_{3\protect\sigma_g}=0.63485797$ a.u., while for the lower panels
$E_{3\protect\sigma_g}=0.63495$ a.u. In both cases, $R=2.068$ a.u.}
\label{fig1}
\end{figure}

In Fig.~\ref{fig1}, we display such results, either computed with a
6-31G gaussian basis set and coefficients obtained from GAMESS-UK
\cite{GAMESS}, or with Slater-type orbitals (\ref{PsiaHFsigm}), and
the coefficients in \cite{Cade66} [upper and lower panels,
respectively]. The outcome of the split-valence computation, displayed in Fig.~%
\ref{fig1}.(a), exhibits a minimum which, for parallel molecular
alignment, is near $\Omega =25\omega $. This is a slightly higher
harmonic order than that observed in \cite{DM2009} (see Fig.~4
therein). The minima observed for the individual $s$ and $p$
contributions, in contrast, agree with the
results presented in \cite{DM2009} (c.f. Fig.~\ref{fig1}.(b) and Fig.~\ref%
{fig1}.(c), respectively). This suggests that the $s$-$p$ mixing possesses
different weights in the present case and in \cite{DM2009}.

The spectra obtained with the Slater-type orbitals, on the other hand, are
practically identical to the results in \cite{DM2009}. This holds both for
the minimum in the full $3\sigma_g$ spectrum [Fig.~\ref{fig1}.(d)], which,
for parallel alignment, is close to $\Omega=21\omega$, and for the patterns
present in the $s$ and $p$ contributions [Fig.~\ref{fig1}.(e) and Fig.~\ref%
{fig1}.(f), respectively]. We have ruled out that this discrepancy
is due to the slightly different ionization potentials employed in
the two computations by performing a direct comparison for the same
set of parameters (not shown). We have also found, employing
GAMESS-UK and several types of basis sets, that the minimum at
$\Omega=25\omega$ is rather robust with respect to small variations
of $E_{3\sigma_g}$ and $R$.
\footnote{%
Apart from the 6-31G basis set mentioned in this paper, which has
been used to compute the spectra in Figs. 1.(a)-(c), we have
employed the following basis sets in GAMESS-UK: STO-3G (Slater-type
orbitals, three Gaussians), and several split-valence basis sets,
namely 3-21G, 4-21G, 4-31G, 5-31G, and 6-21G. For all cases we found
that the two-center interference minimum of the $3\sigma_g$ spectrum
agreed with Fig. 1(a). For more details on split valence basis sets
see, e.g., J. Stephen Binkley, John A. Pople, Warren J. Hehre, J.
Am. Chem. Soc. \textbf{102}, 939 (1980). }. Hence, in comparison to
our computations, it seems that the contributions of the $s$ states
to the spectra are slightly underestimated in \cite{Cade66}.

\begin{figure}[ptb]
\noindent\hspace*{-0.5cm}\includegraphics[width=9.5cm]{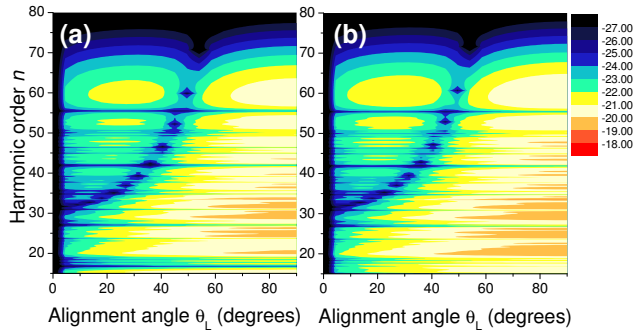}
\caption{High-order harmonic spectra computed using the
$1\protect\pi_{ux}$
(HOMO-1) orbital in $N_2$ as a function of the alignment angle $\protect%
\theta_L$, for the same laser-field parameters as in the previous figure. In
panel (a), we considered a gaussian basis set and computed the coefficients
with GAMESS-UK \protect\cite{GAMESS}, while in panel (b) we took Slater-type
orbitals and the data from \protect\cite{Cade66}. The bound-state energy of
the HOMO and the equilibrium internuclear distance have been taken from the
respective computations. For panel (a), $E_{1\protect\pi%
_u}=0.61544$ a.u., while for panel (b)
$E_{1\protect\pi_u}=0.65087981$ a.u. In both cases, the internuclear
distance is $R=2.068$. Note that in the Slater-type case the
$1\protect\pi_{u}$ orbitals are more loosely bound than the
$3\protect\sigma_g$ orbital.} \label{fig2}
\end{figure}

In Fig.~\ref{fig2}, we present the high-harmonic spectra computed assuming,
instead, that the electron comes back and returns to the $1\pi_{ux}$
orbital, i.e., employing the transition amplitude $M_{22}$. The $1\pi_{uy}$
orbital should behave in a similar way and lead to the same spectrum, as it
exhibits the same dependence with regard to the alignment angle.

As in the $3\sigma_g$ case, we construct the bound-state
wavefunction either from Slater-type orbitals and the data in
\cite{Cade66} or from a 6-31G basis set obtained from GAMESS-UK
\cite{GAMESS}. These results are displayed in Figs.~\ref{fig2}.(a)
and \ref{fig2}.(b), respectively. In both cases, we find that the
two-center interference occurs at the very same harmonic order.
Furthermore, apart from discrepancies in the overall intensity, the
spectra exhibit a very similar substructure. Finally, in both cases,
the yield drops considerably for parallel-aligned molecules. This is
expected, as, if the angle $\theta_L=0$, the $\pi$ orbitals exhibit
a nodal plane along the polarization axis. If the alignment angle
increases, this nodal plane moves further and further away from the
field-polarization axis, and the high-order harmonic yield
increases.

\subsection{Quantum interference of HOMO and HOMO-1}

\label{interference}

We will now investigate which signatures the interference between the $%
3\sigma _{g}$ and $1\pi _{u}$ leave on the high-order harmonic spectra. In
all cases, we will consider both the recombination prefactor $d_{z}^{(\Psi
)}(\mathbf{p+A}(t))$ and the ionization prefactor $d_{z}^{(\Psi )}(\mathbf{%
p+A}(t^{\prime }))$. The ionization prefactor is important in this context
due to the fact that an electron reaching the continuum from a $\sigma $ or
a $\pi$ orbital behaves in very different ways, with regard to the alignment
angle $\theta_L $. In fact, for $\sigma $ orbitals, one expects ionization
to be significant for small $\theta_L $ and to be negligible for large
values of this parameter. For $\pi $ orbitals, due to the presence of the
nodal plane, the opposite behavior is expected to occur.

\subsubsection{Two-dimensional model}

We will commence by addressing the situation for which $\phi _{p}=0$, i.e.,
we are restricting the dynamics of the problem to the $p_{x}p_{z}$ plane. In
this case, the initial wavefunction is a superposition of the $3\sigma _{g}$
and $1\pi _{ux}$ states only, i.e., $C_{1\pi _{uy}}=0$. We consider that it
is equally probable that the electron leaves from each of these states,
i.e., $C_{3\sigma _{g}}=C_{1\pi _{ux}}=1/\sqrt{2}$.

\begin{figure}[tbp]
\noindent\includegraphics[width=9.5cm]{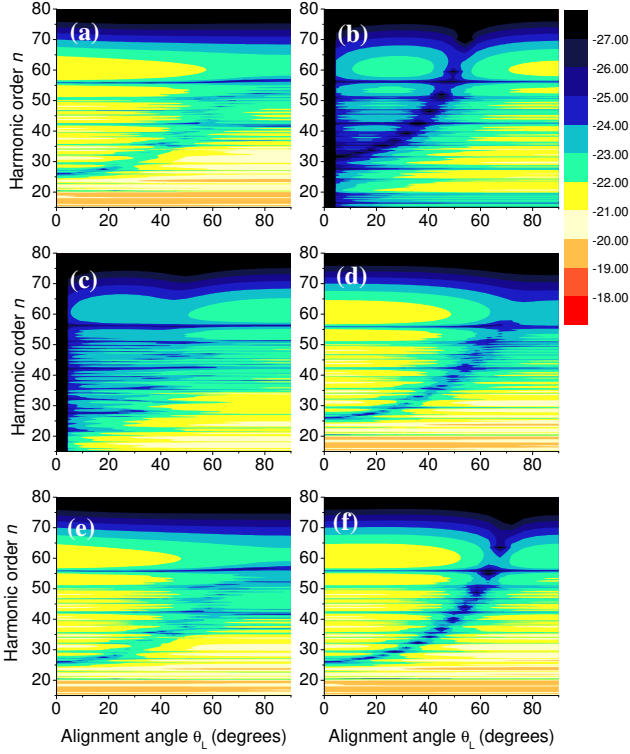}
\caption{Contribution of different processes to the high-harmonic
spectra, as functions of the alignment angle $\protect\theta_L$, for
the same field
parameters in the previous figure. We chose $\protect\phi_p=0$ so that the $1%
\protect\pi_{uy}$ orbital does not contribute. The dipole matrix elements
have been computed using a gaussian basis set and GAMESS-UK \protect\cite%
{GAMESS}. In this case, $E_{3\protect\sigma_g}=0.63485797$ a.u, $E_{1\protect%
\pi_u}=0.65087981$ a.u. and $R=2.068$ a.u. Panel (a): $%
|M_{11}+M_{22}+M_{12}+M_{21}|^{2}$; panel (b): processes finishing at the $1%
\protect\pi_{ux}$ orbital, i.e., $|M_{21}+M_{22}|^{2} $; panel (c):
processes in which the electron starts at one orbital and recombines with
the other, i.e., $|M_{12}+M_{21}|^{2} $; panel (d): processes in which the
electron starts from and returns to the same orbital, i.e., $%
|M_{11}+M_{22}|^2$; panel (e): processes starting at the $3\protect\sigma_g$
orbital, i.e., $|M_{21}+M_{11}|^2$; panel (f): processes finishing at the $3%
\protect\sigma_g$ orbital, i.e., $|M_{12}+M_{11}|^2$.}
\label{fig3}
\end{figure}

In Fig.~\ref{fig3}.(a), we show the full spectrum, in which all the
transition amplitudes \thinspace $M_{j\nu },(j,\nu =1,2)$ are summed
coherently. Especially for small alignment angles, this spectrum exhibits a
minimum very close to that obtained if only the $3\sigma _{g}$ state is
taken. This minimum gets more and more blurred as the alignment angle $%
\theta_L$ increases. Possibly, this is the main influence of the $1\pi _{ux}$
orbital, as its contributions increase with $\theta_L$. In the following, we
will investigate these patterns in more detail. For that purpose, we
consider the quantum interference between specific processes. These results
are depicted in the remaining panels of Fig.~\ref{fig3}.

If the electron recombines with the $1\pi _{ux}$ orbital, regardless of
where it started from [Fig.~\ref{fig3}.(b)], a very pronounced interference
minimum is observed. This minimum occurs for the same harmonic orders as if
only $\pi$ states are taken (c.f. Fig.~\ref{fig2}). This is expected, as
high-order harmonic generation in the former case is only due to
recombination with the $1\pi _{ux}$ orbital, even if two orbitals are
involved.

Apart from that, the yield practically vanishes at $\theta_L=0.$ This
behavior is caused by the nodal plane which exists along the internuclear
axis for the $\pi _{ux}$ orbital. In this case, recombination for both the
transition amplitudes $M_{12}$ and $M_{22}$, and ionization for the
transition amplitude $M_{22},$ are strongly suppressed. For parallel
alignment, this plane is along the laser-field polarization. As the
alignment angle increases, this plane moves away from the field-polarization
axis. Consequently, the yield increases. This explains why, in the overall
spectrum, the minimum is determined by the $3\sigma _{g}$ orbital. For the
parameters considered in this work, such a minimum lies mostly in the region
of small alignment angles.

In contrast, if only processes involving ionization from $1\pi
_{ux}$ and recombination with $3\sigma _{g},$ or vice-versa, are
taken, the double-slit interference minimum is completely blurred
[c.f. Fig. \ref{fig3}.(c)]. This is due to the fact that both
contributions are comparable, and exhibit minima for different
harmonic orders. Furthermore, since either recombination with or
ionization from a $\pi $ state is taking place, a strong suppression
for $\theta _{L}=0$ is present. In Fig. \ref{fig3}, this is the only
case for which we observed a complete disappearance of the
double-slit minimum. Indeed, neither for the processes involving
only one state [\ref{fig3}.(d)], or starting at $3\sigma _{g}$
regardless of the end state [Fig. \ref{fig3}.(e)] does the minimum
completely vanish. However, a sharp minimum is only present if we
take into account the processes in which the
electron recombines with the same state. Concrete examples are Fig.~\ref%
{fig3}.(b), and Fig.~\ref{fig3}.(f), where the processes finishing at $1\pi
_{ux}$ and $3\sigma _{g}$, respectively, are presented. In Fig.~\ref{fig3}%
.(f), we also notice that, for $0\leq \theta _{L}\leq 45^{\circ }$,
the contributions from the $\sigma $ orbital are up to two orders of
magnitude larger than those from the $\pi $ orbital [i.e.,
Fig.~\ref{fig3}.(b)]. This is further evidence that the minimum is
determined by the $3\sigma_g $ state.

\subsubsection{Three-dimensional case}

In a more realistic situation, one cannot restrict the electron
dynamics only to the $p_{x}p_{z}$ plane. In fact, there exist two
$\pi $ orbitals which, even though they behave in the same way with
respect to the alignment angle $\theta _{L},$ are degenerate. Hence,
they provide a completely different weight to the states $\left\vert
\psi _{0}\right\rangle$ from which the electron is released and to
which it returns. Furthermore, under many experimental conditions,
the azimuthal angle $\phi _{p}$ cannot be resolved. Thus, this
parameter must be integrated over.

Explicitly, the resulting spectrum is given by%
\begin{equation}
S(\Omega )=\int_{0}^{2\pi }|\sum\limits_{j,\nu }M_{j\nu }|^{2}d\phi _{p}.
\label{sum}
\end{equation}%
For the specific problem addressed in this work, the above-stated
sum consists of 81 terms. In general, the integrand in Eq
(\ref{sum}) is of the form $M_{\alpha \beta }^{\ast }M_{j\nu }.$ Its
general dependence on the azimuthal angle $\phi _{p}$ is given by
$(\sin \phi _{p})^{\eta _{1}}(\cos \phi _{p})^{\eta _{2}},$ where
the exponents $\eta _{1},\eta _{2}$ are integers. Depending on such
exponents, the contributions to the full harmonic spectrum carry
different weights. For $\eta _{1}$ or $\eta _{2}$ odd, the
contributions to the spectrum vanish. This implies that only the
terms $M_{jj}^{\ast }M_{\nu \nu },$ for any $j,\nu ,$ and $M_{\nu
j}^{\ast }M_{j\nu },$ $M_{j\nu }^{\ast }M_{j\nu },$ for $j\neq \nu
,$ survive.

If the integral in (\ref{sum}) is carried out, one obtains%
\begin{eqnarray}
S(\Omega ) &\sim &\sum\limits_{\alpha ,\beta,\nu ,j}\mathcal{W}(\alpha ,\beta,\nu ,j)%
\mathcal{M}_{\alpha \beta }^{\ast }(p+A(t),\theta _{p})  \notag \\
&&\times \mathcal{M}_{j\nu }(p+A(t^{\prime }),\theta _{p}),
\end{eqnarray}%
where $\mathcal{M}_{j\nu }(p+A(\tau ),\theta _{p})$, $\tau
=t,t^{\prime }$ is the transition amplitude without the dependence
on $\phi _{p}.$ In Table 1, we provide it the weights
$\mathcal{W}(\alpha ,\beta ,\nu ,j)$ for each term in the sum
(\ref{sum}), after integration over $\phi _{p}.$

\begin{center}
\begin{table}[tbp]
\vspace*{0.5cm} $%
\begin{tabular}{cccccccccc}
\hline\hline
& $M_{11}$ & $M_{12}$ & $M_{13}$ & $M_{21}$ & $M_{22}$ & $M_{23}$ & $M_{31}$
& $M_{32}$ & $M_{33}$ \\ \hline
$M_{11}$ \vline & $2\pi $ & 0 & 0 & 0 & $\pi $ & 0 & 0 & 0 & $\pi $ \\
$M_{12}$ \vline & 0 & $\pi $ & 0 & $\pi $ & 0 & 0 & 0 & 0 & 0 \\
$M_{13}$ \vline & 0 & 0 & $\pi $ & 0 & 0 & 0 & $\pi $ & 0 & 0 \\
$M_{21}$ \vline & 0 & $\pi $ & 0 & $\pi $ & 0 & 0 & 0 & 0 & 0 \\
$M_{22}$ \vline & $\pi $ & 0 & 0 & 0 & $3\pi /4$ & 0 & 0 & 0 & $\pi /4$ \\
$M_{23}$ \vline & 0 & 0 & 0 & 0 & 0 & $\pi /4$ & 0 & $\pi /4$ & 0 \\
$M_{31}$ \vline & 0 & 0 & $\pi $ & 0 & 0 & 0 & $\pi $ & 0 & 0 \\
$M_{32}$ \vline & 0 & 0 & 0 & 0 & 0 & $\pi /4$ & 0 & $\pi /4$ & 0 \\
$M_{33}$ \vline & $\pi $ & 0 & 0 & 0 & $\pi /4$ & 0 & 0 & 0 & $3\pi /4$ \\
\hline\hline
\end{tabular}%
$%
\caption{Weights $\mathcal{W}(\protect\alpha
,\protect\beta,\protect\nu ,j)$ for the
contributions of the terms $M_{\protect\alpha \protect\beta }^{\ast }M_{j%
\protect\nu}$ to the spectra, when integrated over the azimuthal angle $%
\protect\phi_p$.}
\end{table}
\end{center}

\begin{figure}[tbp]
\noindent\hspace*{-0.5cm}\includegraphics[width=10cm]{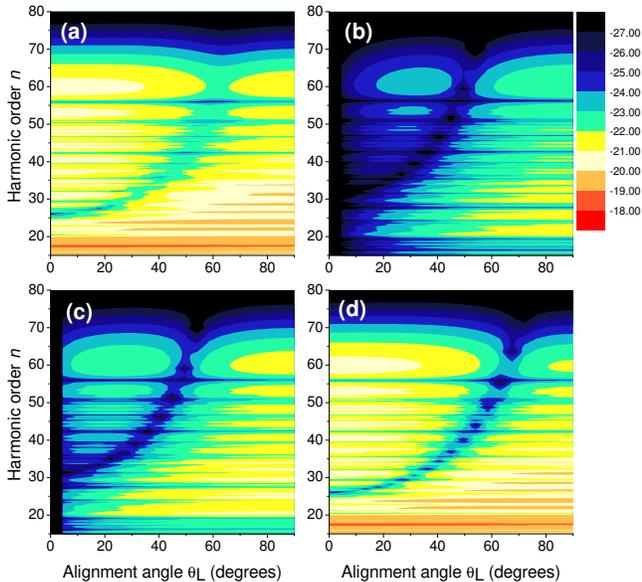}
\caption{Contribution of different processes to the high-harmonic
spectra, as functions of the alignment angle $\protect\theta_L$, for
the same field parameters in the previous figure, taking into
account the degeneracy of the $1\protect\pi_u$ orbitals in a
three-dimensional scenario, and non-resolved angle $\protect\phi_p$.
The dipole matrix elements have been computed using
a gaussian basis set and GAMESS-UK \protect\cite{GAMESS}. In this case, $E_{3%
\protect\sigma_g}=0.63485797$ a.u, $E_{1\protect\pi_u}=0.65087981$ a.u. and $%
R=2.068$ a.u. Panel (a): all processes,i.e., the full sum in Eq.
\protect{\ref{sum}}; panel (b): processes incorporating the
$1\protect\pi_{u}$ orbitals only, i.e.,
$|M_{22}+M_{23}+M_{32}+M_{33}|^2$ in \protect{\ref{sum}}; panel (c):
processes in which the electron starts at any orbital and recombines
with the $1\pi_u$ orbitals,
i.e.,$|M_{22}+M_{23}+M_{32}+M_{33}+M_{21}+M_{31}|^2$; panel (d):
processes in which the electron recombines with the $3\sigma_g$
orbital, i.e., $|M_{11}+M_{12}+M_{13}|^2$.} \label{fig4}
\end{figure}

In Fig.~\ref{fig4}, we depict the high-order harmonic spectra
obtained employing Eq. (\ref{sum}), starting by the full spectrum
[Fig.~\ref{fig4}.(a)]. Therein, the minimum caused by the
recombination of the electron with the $3\sigma_g$ orbital is
clearly visible, and the blurring due to the influence of the
degenerate $1\pi_u$ orbitals is even less pronounced than for its
two-dimensional counterpart. At first sight, this is a
counterintuitive finding, as, in the three-dimensional case, there
are many more processes involving the latter orbitals. Possibly,
this is a consequence of two main effects. First, due to the
presence of the nodal plane, for a broad range of alignment angles
the contributions of the $\pi$ orbitals are strongly suppressed.
Second, in general, the weights $\mathcal{W}(\alpha ,\beta,\nu ,j)$
involving the $3\sigma_g$ orbitals are larger than those involving
the $1\pi_u$ orbitals only. This is, once more, a consequence of the
geometry of the latter orbitals.

In order to investigate this fact, we computed the high-order
harmonic spectrum taking into account only the latter contributions.
Such results are displayed in Fig.~\ref{fig4}.(b). Qualitatively,
the spectrum obtained in this way is in perfect agreement with those
displayed in Fig.~\ref{fig2}, which have been computed using
$1\pi_{ux}$ only, or with that shown in Fig.~\ref{fig3}.(b), which
incorporates the processes in which the electron recombines at
$1\pi_{ux}$ in a two-dimensional scenario. In fact, all such spectra
exhibit a minimum above $\Omega=31\omega$ for $\theta_p$ in the
vicinity of zero, which moves towards the cutoff for
$\theta_L=45^{\circ }$. The harmonic yield in Fig.~\ref{fig4}.(b),
especially in the region of small alignment angles, is up to the
three orders of magnitude weaker than the full spectrum. This huge
discrepancy, however, would not cause any blurring in the full
spectrum.

Potentially, the blurring may also be caused by the processes in
which the electron is released from the $3\sigma_g$ orbital and
recombines with \emph{any} of the $1\pi_{u}$ orbitals. In this
latter case, a minimum  near $\Omega=31\omega$ for $\theta_L$ in the
vicinity of zero would also be present. Therefore, such processes
must be incorporated. In Fig.~\ref{fig4}.(c), we consider the
contributions from all possible processes finishing at the
$1\pi_{u}$ orbitals, regardless of where the electron left from. As
expected, there is a substantial increase in the yield for small
angles, in comparison to Fig.~\ref{fig4}.(b).

Such increase is however not sufficient to match the contributions
from the $3\sigma_g$ states to the full spectrum. In
Fig.~\ref{fig4}.(d), we show that, for small alignment angles, the
processes for which the electron recombines with the $3\sigma_g$
state dominate. In fact, for $0<\theta_L<45^{\circ}$ the yield in
Fig.~\ref{fig4}.(d) is roughly one order of magnitude larger than
that displayed in Fig.~\ref{fig4}.(c). This is not obvious, as there
are twice as many more processes contributing to the yield in this
latter case, namely six against three. A direct comparison of
Figs.~\ref{fig4}.(a) and ~\ref{fig4}.(d) also shows the
above-mentioned dominance for small angles. For larger angles, the
contributions from the $1\pi_u$ orbitals start to play a more
significant role and there is an increase in the blurring. In all
cases for which the electron only starts from or recombines with the
$1\pi_u$ orbitals [Figs.~\ref{fig4}.(b) and \ref{fig4}.(c)], there
is a strong suppression of the yield for parallel alignment. This is
due to the fact that the nodal plane along the molecular axis
coincides with the laser-polarization axis in this case.
\section{Conclusions}
\label{concl}

We considered the influence of two closely lying molecular orbitals
on the high-order harmonic spectrum from $N_2$: the $1\pi_u$ and
$3\sigma_g$ orbitals. We employed a very simple model, in which the
strong-field approximation has been modified in order to incorporate
the situations in which an electron leaves from the $3\sigma_g$
orbital and recombines with $1\pi_u$ and vice-versa. We have also
included the degeneracy of the $1\pi_u$ orbital. We made a detailed
assessment of the contributions of all possible processes to the
high-order harmonic spectra.

The main conclusion to be drawn from this work is that the shape and
the two-center interference patterns observed for the high-order
harmonic spectra from $N_2$ are mainly determined by the $3\sigma_g$
orbital, even though the $1\pi_u$ orbitals are energetically very
close. The main effect of the latter orbitals is to introduce some
blurring in the interference minimum determined by $3\sigma_g$.

Physically, this is due to the particular geometry of the $1\pi_u$
orbitals. Indeed, for small alignment angles $\theta_L$, these
orbitals exhibit a nodal plane close to the polarization axis, so
that tunneling and recombination are strongly suppressed. Hence, in
this region, the high-order harmonic spectra are mainly dominated by
the $3\sigma_g$ orbital. We have verified that this dominance
extends up to approximately $\theta_L=45^{\circ}$. For the
parameters considered in this paper, the two-center minimum occurs
within this region, so that it is mainly determined by the
$3\sigma_g$ state.

Furthermore, due to their nontrivial dependence on the azimuthal
angle, the $1\pi_u$ orbitals carry less weight when this parameter
is integrated over. Interestingly, even if a three-dimensional
computation is carried out and the degeneracy of the $1\pi_u$
orbitals is considered, this angular dependence outweighs the fact
that there are more processes in which the electron recombines with
one of the $1\pi_u$ orbitals.

We have also shown that, due to the above-mentioned non-trivial
angular dependence, the influence of such orbitals is over-estimated
if the dynamics of the problem is reduced to the $p_xp_z$ plane,
i.e., if the azimuthal angle $\phi_p$ is chosen to be vanishing.
Such an approximation has been extensively used in the literature
(see, e.g., \cite{DM2009} in with HHG from the $\pi_g$ orbital of
the $O_2$ molecule has been computed). This is not an obvious
result, as two-dimensional models do not consider the degeneracy of
the $1\pi_u$ orbital.

Finally, it is worth mentioning that the findings of this paper
agree qualitatively with recent results obtained employing more
sophisticated methods, such as Dyson orbitals and many-body
perturbation theory \cite{Multielectron}. Therein, it has been shown
that many-electron effects did not play a significant role in the
bound-state reconstruction of $N_2$, and that the information
retrieved from the spectra was mostly related to the $3\sigma_g$
orbital.

It may, however, be possible to identify the influence of the
$1\pi_u$ orbitals by looking at effects for perpendicular-aligned
molecules, or relatively large alignment angles. In this case, the
contributions from  $3\sigma_g$ to the high-order harmonic spectrum
are not expected to obfuscate those from $1\pi_u$. In fact,
recently, the influence of the latter orbitals on the HHG spectrum
of $N_2$ has been identified experimentally for
perpendicular-aligned molecules, in form of a maximum at the
rotational half-revival \cite{McFar2008}.

\acknowledgements We would like to thank D. B. Milo\v{s}evi\'{c}, J.
Tennyson, R. Torres, H. J. J. van Dam and P. Durham for useful
discussions, and M. T. Nygren for his collaboration in the early
stages of this project. We are particularly indebted to H. J. J. van
Dam for his help with GAMESS. We are also grateful to the Daresbury
laboratory and the ICFO-Barcelona for their kind hospitality. This
work has been financed in part by the UK EPSRC (Grant no.
EP/D07309X/1).


\begin{thebibliography}{99}
\bibitem{Itatani} J. Itatani, J. Levesque, D. Zeidler, H. Niikura, H. P\'{e}%
pin, J. C. Kieffer, P. B. Corkum and D. M. Villeneuve, Nature
\textbf{432}, 867 (2004); W. Boutu, S. Haessler, H. Merdji, P.
Breger, G. Waters, M. Stankiewicz, L. J. Frasinski, R. Taieb, J.
Caillat, A. Maquet, P. Monchicourt, B. Carr\'{e} and P.
Sali\`{e}res, Nature Physics \textbf{4}, 545 (2008).

\bibitem{attomol} H. Niikura, F. L\'{e}gar\'{e}, R. Hasbani, A. D. Bandrauk,
M. Yu. Ivanov, D. M. Villeneuve and P. B. Corkum, Nature
\textbf{417}, 917 (2002); \ H. Niikura, F. L\'{e}gar\'{e}, R.
Hasbani, M. Yu. Ivanov, D. M. Villeneuve and P. B. Corkum, Nature
\textbf{421}, 826 (2003); S. Baker, J. S. Robinson, C. A. Haworth,
H. Teng, R. A. Smith, C. C. Chiril\u{a}, M. Lein, J. W. G. Tisch, J.
\ P. Marangos, \ Science \textbf{312}, 424 (2006).

\bibitem{interfexp} B. Shan, X. M. Tong, Z. Zhao, Z. Chang, and C. D. Lin,
Phys. Rev. A \textbf{66}, 061401(R) (2002); F. Grasbon, G. G.
Paulus, S. L. Chin, H. Walther, J. Muth-B\"{o}hm, A. Becker and F.
H. M. Faisal, Phys. Rev. A \textbf{63}, 041402(R)(2001); C. Altucci,
R. Velotta, J. P. Marangos, E. Heesel, E. Springate, M. Pascolini,
L. Poletto, P. Villoresi, C. Vozzi, G. Sansone, M. Anscombe, J. P.
Caumes, S. Stagira, and M. Nisoli, Phys. Rev. A \textbf{71}, 013409
(2005); T. Kanai, S. Minemoto and H. Sakai, Nature \textbf{435}, 470
(2005).

\bibitem{tstep} P. B. Corkum, Phys. Rev. Lett. \textbf{71}, 1994 (1993); K.
C. Kulander, K. J. Schafer, and J. L. Krause in: B. Piraux et al.
eds., \emph{Proceedings of the SILAP conference}, (Plenum, New York,
1993).

\bibitem{doubleslit} M. Lein, N. Hay, R. Velotta, J. P. Marangos, and P. L.
Knight, Phys. Rev. Lett. \textbf{88}, 183903 (2002); Phys. Rev. A \textbf{66}%
, 023805 (2002); M. Spanner, O. Smirnova, P. B. Corkum and M. Y.
Ivanov, J. Phys. B\textbf{\ 37}, L243 (2004).

\bibitem{MoreN2}R. Torres and J. P. Marangos, J. Mod. Opt.
\textbf{54}, 1883 (2007); M. G\"{u}hr, B. K. McFarland, J. P. Farrel
and P. H. Bucksbaum, J. Phys. B \textbf{40}, 3745 (2007).

\bibitem{MadsenN2} See, e.g., C. B. Madsen and L.B. Madsen, Phys. Rev. A
\textbf{74}, 023403 (2006); C. B. Madsen, A. S. Mouritzen, T. K. Kjeldsen,
L. B. Madsen, Phys. Rev. A \textbf{76}, 035401 (2007).

\bibitem{DM2009} S. Od\v{z}ak and D. B. Milo\v{s}evi\'{c}, Phys. Rev. A
\textbf{79}, 023414 (2009); J. Phys. B \textbf{42}, 071001 (2009).

\bibitem{UsachenkoN2ion} V. I. Usachenko, and S. I. Chu, Phys. Rev. A
\textbf{71}, 063410 (2005); Vladimir I. Usachenko, Phys. Rev. A \textbf{73},
047402 (2006); V. I. Usachenko, P. E. Pyak, and Shih-I Chu, Laser Phys. \
\textbf{16}, 1326 (2006); Vladimir I. Usachenko, Pavel E. Pyak, and
Vyacheslav V. Kim, Phys. Rev. A 79, 023415 (2009).

\bibitem{DM2006} D. B. Milo\v{s}evi\'{c}, Phys. Rev. A \textbf{74}, 063404
(2006); M. Busulad\v{z}i\'{c}, A.
Gazibegovi\'{c}-Busulad\v{z}i\'{c},D. B. Milo\v{s}evi\'{c}, and W.
Becker, Phys. Rev. Lett. \textbf{100}, 203003 (2008); Phys. Rev. A
\textbf{78}, 033412 (2008).

\bibitem{LinN2}A. T. Le, R. R. Lucchese, S. Tonzani, T. Morishita, and
C. D. Lin, Phys. Rev. A \textbf{80}, 013401 (2009).

\bibitem{McFar2008}Brian K. McFarland, Joseph P. Farrell, Philip H. Bucksbaum, Markus
G\"{u}hr, Science \textbf{322} (5905), 1194 (2008).

\bibitem{Multielectron} S.
Patchkovskii, Z. Zhao, T. Brabec and D. M. Villeneuve, Phys. Rev.
Lett. \textbf{97}, 123003 (2006); S. Patchkovskii, Z. Zhao, T.
Brabec, and D.M. Villeneuve, J. Chem. Phys. \textbf{126}, 114306
(2007); O. Smirnova, S. Patchkovskii, Y. Mairesse, N. Dudovich, D.
Villeneuve, P. Corkum, and M. Yu. Ivanov, Phys. Rev. Lett.
\textbf{102}, 063601 (2009).

\bibitem{degeneracy} C.B. Madsen and L.B. Madsen, Phys Rev. A \textbf{76},
043419 (2007).


\bibitem{gauge} C. C. Chiril\u{a} and M. Lein, Phys. Rev. A \textbf{73},
023410 (2006).

\bibitem{dressedSFA} O. Smirnova, M. Spanner and M. Ivanov, J. Phys. B
\textbf{39}, S307 (2006).

\bibitem{JMOCL2006} C. C. Chiril\u{a} and M. Lein, J. Mod. Opt. \textbf{54},
1039 (2007); G. N. Gibson and J. Biegert, Phys. Rev. A \textbf{78},
033423 (2008).


\bibitem{F2007} C. Figueira de Morisson Faria, Phys. Rev. A \textbf{76},
043407 (2007).

\bibitem{SSY2007} O. Smirnova, M. Spanner and M. Ivanov, J. Mod. Opt.
\textbf{54}, 1019 (2007).

\bibitem{hhgsfa} M. Lewenstein, Ph. Balcou, M. Yu. Ivanov, A. L'Huillier and
P. B. Corkum, Phys. Rev. A \textbf{49}, 2117 (1994);W. Becker, A. Lohr, M.
Kleber, and M. Lewenstein, Phys. Rev. A \textbf{56}, 645 (1997).

\bibitem{orbitshhg} P. Sali\`{e}res, B. Carr\'{e}, L. LeD\'{e}roff, F.
Grasbon, G. G. Paulus, H. Walther, R. Kopold, W. Becker, D. B. Milo\v{s}evi%
\'{c}, A. Sanpera and M. Lewenstein, Science \textbf{292}, 902
(2001).

\bibitem{atiuni} C. Figueira de Morisson Faria, H. Schomerus and W. Becker,
Phys. Rev. A \textbf{66}, 043413 (2002).

\bibitem{singularity} C. Figueira de Morisson Faria and M. Lewenstein, J.
Phys. B \textbf{38}, 3251 (2005).

\bibitem{Cade66} P. E. Cade, K. D. Sales and A. C. Wahl, J. Chem. Phys.
\textbf{44}, 1973 (1966).

\bibitem{GAMESS} GAMESS-UK is a package of ab initio programs. See:
"http://www.cfs.dl.ac.uk/gamess-uk/index.shtml", M.F. Guest, I. J. Bush,
H.J.J. van Dam, P. Sherwood, J.M.H. Thomas, J.H. van Lenthe, R.W.A Havenith,
J. Kendrick, Mol. Phys. \textbf{103}, 719 (2005).
\end{thebibliography}
\end{document}